\documentclass[apl,aps,singlecolumn,superscriptaddress]{revtex4-1}
\usepackage{graphics}
\usepackage{epsf}
\usepackage{amsmath}
\usepackage{amsfonts}
\usepackage{amssymb}
\usepackage{graphicx}
\usepackage{epsfig}
\usepackage{color}
\newcommand{\beq}{\begin{equation}}
\newcommand{\eeq}{\end{equation}}

\begin{document}
\title{Polarity compensation in ultra-thin films of complex oxides: The case of a perovskite nickelate}
\author{S. Middey}
\email[Electronic address: ]{smiddey@uark.edu  }
\affiliation{Department of Physics, University of Arkansas, Fayetteville, Arkansas 72701, USA}
\author{P. Rivero}
\affiliation{Department of Physics, University of Arkansas, Fayetteville, Arkansas 72701, USA}
\author{D. Meyers}
\affiliation{Department of Physics, University of Arkansas, Fayetteville, Arkansas 72701, USA}
\author{M. Kareev}
\affiliation{Department of Physics, University of Arkansas, Fayetteville, Arkansas 72701, USA}
\author{X. Liu}
\affiliation{Department of Physics, University of Arkansas, Fayetteville, Arkansas 72701, USA}
\author{Y. Cao}
\affiliation{Department of Physics, University of Arkansas, Fayetteville, Arkansas 72701, USA}
\author{J. W. Freeland}
\affiliation{Advanced Photon Source, Argonne National Laboratory, Argonne, Illinois 60439, USA}
\author{S. Barraza-Lopez}
\affiliation{Department of Physics, University of Arkansas, Fayetteville, Arkansas 72701, USA}
\author{ J. Chakhalian}
\affiliation{Department of Physics, University of Arkansas, Fayetteville, Arkansas 72701, USA}

\begin{abstract}
We address the fundamental issue of growth of perovskite ultra-thin films under the condition of a strong polar mismatch at the heterointerface exemplified by the growth of a correlated metal LaNiO$_3$ on the band insulator SrTiO$_3$ along the pseudo cubic [111] direction. While in general the metallic LaNiO$_3$ film can effectively screen this polarity mismatch, we establish that in the ultra-thin limit, films are insulating in nature and require  additional chemical and structural reconstruction  to compensate for such mismatch. A  combination of in-situ reflection high-energy electron diffraction  recorded during the growth, X-ray diffraction, and synchrotron based resonant X-ray spectroscopy reveal the formation of a chemical phase La$_2$Ni$_2$O$_5$ (Ni$^{2+}$) for a few unit-cell thick films. First-principles layer-resolved calculations of the potential energy across the nominal LaNiO$_3$/SrTiO$_3$ interface confirm that the oxygen vacancies can efficiently reduce the electric field at the interface.  

\end{abstract}

\maketitle


Artificial heterostructures based on transition metal oxides lead to fascinating interfacial phenomena unfeasible in their bulk form~\cite{jak_nm,tokura_nm_12,ramesh_mat,bibes}. For example, highly mobile two dimensional electron gas~\cite{lao_nature} and   superconductivity around 200 mK ~\cite{lao_science} have been observed at the interface between  two wide band gap insulators LaAlO$_3$ (LAO) and SrTiO$_3$ (STO). Though the actual physical process responsible for this metallic behavior is still under debate, it has been unanimously concluded that the polar mismatch  at the LAO/STO interface is responsible for this  emergent metallic state~\cite{lao_nm,lao_sto_opt,lao_sto_hardxps,lao_sto_mixing,lao_sto_vacancy1,lao_sto_vacancy,lao_admi}. With LAO/STO as a non-trivial example, the polar mismatch driven phenomena is now of great general interest and has been studied in a broader range of complex oxide materials~\cite{llvo_sto,lao_lvo,lco_sto,lufeo3}.  The extensive experimental work has demonstrated that the divergence of the electric field due to this interfacial polarity mismatch can be resolved in a number of ways, $e.g.$ by electronic reconstruction~\cite{llvo_sto,lao_lvo,lao_nm,lao_sto_opt,lao_sto_hardxps}, by cationic inter-mixing~\cite{lco_sto,lao_sto_mixing}, or by oxygen vacancies~\cite{lao_sto_vacancy1,lao_sto_vacancy_prb,lao_sto_vacancy}.

Furthermore, recent theoretical works  have  predicted a host of exotic ground states ($e.g.$ Dirac half-semimetals, quantum-anomalous Hall-insulators, ferromagnetic nematics, time-reversal invariant Z$_2$ topological phases, Chern insulator, etc.) in heterostructures of transition metal oxides grown along the pseudo-cubic [111] direction~\cite{lno1_fiete,satoshi_prb,sriro_prb,nagaosa_nc,lno_fiete,okamoto_prl,lno_strain_fiete}. To this  end, the common substrate SrTiO$_3$ (or LaAlO$_3$, NdGaO$_3$, YAlO$_3$, LaGaO$_3$, etc.) forms alternating +4e,  $-$4e   (or +3e, $-$3e) charged planes stacked along this direction, leading to a large polarity mismatch at the interface (see Fig.~1(b)). This mismatch can have a significant effect on initial nucleation and growth~\cite{blok} and on the overall materials properties that  are particularly sensitive for ultra-thin films. Based on this,  the mechanism by which polar discontinuity is compensated in real materials is  of  paramount interest towards the experimental realization of new materials with exotic properties.

To address this  issue, rare earth nickelate LaNiO$_3$ (LNO) films were grown on a single crystal cubic STO substrate oriented in the strongly polar [111] direction. In bulk LNO, Ni is in the low-spin 3$d^7$ configuration with an unusually high +3 oxidation state, i.e. ($t_{2g}^6$, $e_g^1$); the compound  is metallic, and it remains paramagnetic down to the lowest temperature. Ultra-thin films of LNO grown along the moderately polar (001) direction have been extensively investigated recently~\cite{jian_apl,lno_jian_prb,lno_epl,lno_keimer_science,lno_keimer_nm,lno_triscone,jak_prl}  largely due to  the prediction of high-temperature superconductivity when heterostructured with the band insulator LAO~\cite{lno_th1,lno_th2}.
The experimentally realized LNO films and LNO/LAO superlattices on STO (001), however, revealed the presence of an unexpected  transition to the Mott insulating and antiferromagnetic ground state with decreasing LNO thickness~\cite{lno_jian_prb,lno_keimer_science,lno_triscone}. Additionally, it was established~\cite{jian_apl} that growth  of the properly  stoichiometric 1uc LNO/ 1uc LAO (uc = unit cell) superlattices requires an extra `buffer' monolayer of LAO to compensate for the polar mismatch, with resonant  x-ray absorption on Ni  $L_{3,2}$-edge confirming the presence of Ni$^{2+}$ in the first LNO layer~\cite{jian_apl}.
%


A direct  inspection of the difference  in the ionic arrangement  of (001) and (111) planes shown in Fig. 1(a) and (b) implies that the expected dipole electric field  mismatch is markedly more severe for ultra-thin LNO films grown along the  (111) orientation and likely requires further compensation mechanisms beyond those already observed on (001) STO. Obviously, when the LNO layer becomes thicker  and  reestablishes its bulk-like metallicity, the polar catastrophe can be easily avoided by metallic screening, without the requirement of additional chemical, structural, or electronic reconstructions~\cite{blok}.
Here,  by  comprehensive experimentation combined with $ab$ $initio$ calculations, we demonstrate how the  growth of La$_2$Ni(2+)$_2$O$_5$ (LNO225) phase from the removal  of oxygen from the octahedral units near the STO/LNO interface can effectively circumvent  such polar catastrophe and leads to a recovery of  stoichiometric layer-by-layer growth despite the highly  polar interface.


Fig. 1(c) shows the set of   RHEED (reflection high energy electron diffraction) patterns recorded from bare STO (111) substrate and after  each consecutive unit cell (uc) during the growth of 10 pseudo-cubic unit cells of  LNO. First of all, the absence of any  additional reflections between specular and off-specular spots in RHEED  pattern of STO (111) substrate~\cite{sto111} (the diffraction pattern recorded under 5x10$^{-7}$ Torr pressure at the growth temperature   has been shown in  Supplemental Figure S1 for better resolution) indicates the absence of  surface reconstructions with the growth conditions used in present study.
It can be clearly seen that during the initial phase of growth (from 1 uc to 5 uc), the RHEED pattern consists of additional Bragg reflections (highlighted by the dotted ellipses) along with the conventional specular and off-specular reflections. This  observation signals the growth of additional epitaxial 3D microcrystallites~\cite{rheed1}, resulting in a rough, phase impure surface.  As growth continues beyond 5 uc, however, the extra spots suddenly merge  and the  new pattern  attains the streak-like  specular (and off-specular) pattern of reflections  common for electron diffraction off an atomically flat 2D surface~\cite{misha_jap,edoped}. Direct space imaging by AFM (atomic force microscope), shown in Fig. 1(d),  confirms the high morphological quality of the grown surface (average surface roughness $\sim$ 180 pm). To investigate the phases within the initial layers, growth was interrupted after reaching 5 uc, and the thin film was annealed following the same protocol used for the 10 uc film. The retention of the additional reflection in the RHEED image even after annealing (Fig.~1(e)), and an AFM  scan (not shown) with roughness $\sim$  480 pm [much higher compared to the roughness of 10 uc film  (180 pm)] indicate that annealing can not improve the surface morphology of these thin-films.
To investigate whether the observed drastic change in the growth phase is related to  the polarity issue, we have performed identical growth on a LAO (111) substrate [\textit{no-polar mismatch} at the LNO/LAO (111) interface]. As seen in  Fig. 1(f) (see  Ref. ~\cite{own_apl} for additional information) the additional reflections are entirely  absent for a 5 uc LNO film grown on the LAO (111) substrate. This result (together with Fig. 1(e) for 5 uc LNO on STO (111))  suggests that the unexpected initial  growth behavior on STO (111) is probably due to the polar discontinuity at the heterointerface.

To gain further insight into the structural quality and the chemical phase  obtained during  the initial growth sequence, we  recorded X-ray diffraction 2$\theta$ - $\omega$ patterns using Cu K$_\alpha$ radiation leading to several key observations. First, as seen in Fig.~2(a), the film  diffraction peak (indicated by arrows) obtained at higher 2$\theta$ than the STO (111)  reflection  confirms the epitaxial [111] growth under tensile strain. The relative width of this peak decreases in moving from 5 uc to 15 uc, due to the increase of film thickness. In  addition, for the 5 uc film, an additional broad peak marked by the $\ast$ has been observed around 2$\theta$ = 35$^\circ$ (see Fig 2(a)).
The presence of oxygen vacancy, discussed latter with absorption data, further highlights that this additional peak in XRD  must originate from some oxygen deficient perovskites  with formulae La$_n$Ni$_n$O$_{3n-1}$. This additional phase present in the 5 uc film has a set of parallel planes with a separation of $d=$2.59 \AA\, which is very close to $d$ of (202) planes for the monoclinic La$_2$Ni$_2$O$_5$ ($d =$ 2.609 \AA \ ) ~\cite{la2ni2} and also for (011) plane of tetragonal LaNiO$_2$ ($d =$ 2.57 \AA \ )~\cite{lani} . The presence of large number of Ni in the +2 oxidation state there (discussed later with Fig. 2(b)) indeed confirms the formation of La$_2$Ni$_2$O$_5$Ê and such a small deviation of the inter-plane separation from the corresponding value of bulk La$_2$Ni$_2$O$_5$ likely arises from the substrate-induced strain.Ê The origin of the gradual shifting of this peak  toward higher angles as the film thickness increases above 5 uc and it's absence in the 15 uc film will be discussed latter. As  x-rays can penetrate deeper compared with the electron beam, the interfacial additional phase can be observed in 10 uc films, while RHEED, being a surface sensitive tool, shows a flat 2D surface (Fig. 1(c)).

%

To  corroborate structural information  from XRD and RHEED with the  knowledge of the electronic structure  and charge state of Ni, we performed resonant X-ray  absorption measurements on Ni $L_{3,2}$-edge and  oxygen $K$-edge in the bulk-sensitive total fluorescence yield (TFY) mode. From the chemistry perspective, the presence of oxygen vacancies should alter the Ni$^{+3}$ charge state. Fig. 2(b) shows the results of the experiment. Due to the strong overlap of the La $M_4$-edge with the Ni $L_3$-edge (Supplemental Figure S5), only Ni $L_2$-edge spectra are used for the analysis, along with the reference spectra from the bulk Ni(2+)O and LaNi(3+)O$_3$ compounds.
 A comparison of  the line-shape  and the $L_2$ energy position (i.e. chemical  shift) for the  5 uc film to the well known Ni$^{2+}$ material NiO  unequivocally indicates that Ni ions are indeed in the +2 oxidation state,  thus providing a direct spectroscopic signature for the existence of the La$_2$Ni$_2$O$_5$ phase as elucidated from the XRD data. As seen in Fig. 2(b) as the film thickness increases the relative weight of Ni$^{+3}$ ions progressively increases as well, causing the relative intensity of the peak near 870 eV to decrease; the absorption spectrum for the 15 uc film is characteristic of Ni$^{3+}$ ions as confirmed by comparison to the  reference  LaNi(3+)O$_3$ absorption.
To verify if the observed Ni$^{2+}$ state is specific  to the interface with  polar discontinuity, we repeated the same XAS measurement on an identical 5 uc LNO film grown on LAO (111) (i.e. no polar jump at the interface). As shown in Fig. 2(b), in this case the 870 eV peak characteristic of Ni$^{2+}$ is absent and the resulting line shape and the absorption peak maximum is consistent  with the  bulk like Ni$^{3+}$ charge state.  This spectroscopic result is in a good  agreement with our diffraction data and lends strong  support to  the notion that the polarity mismatch at the LNO-STO interface stabilizes the La$_2$Ni$^{2+}_2$O$_5$ phase in the ultra thin limit.

The presence of oxygen vacancies not only alters local charge states of nickel but also implies a difference in hybridization of the Ni-O bond.
The effect of oxygen vacancies was investigated by performing absorption measurements on the O $K$-edge and monitoring the pre-edge region around 529 eV. This pre-edge intensity originates from a 3$d^8$\underline{$L$} $\rightarrow$  \underline{$c$}3$d^8$ transition~\cite{oxygen_kedge}, where \underline{$L$} and \underline{$c$} denotes a ligand hole and an O 1$s$ core hole, respectively. In the past, the reduction of the O K pre-edge intensity was used to monitor the oxygen vacancy  formation in purposely reduced bulk compounds of  LaNiO$_{3-\delta}$~\cite{oxygen_kedge}. As shown in Fig. 2(c),  the intensity of the pre-peak around 528.4 eV for the 15 uc film is very similar to that of the stoichiometric LaNiO$_3$ and as the film thickness falls below 6 uc the pre-peak intensity disappears, implying dramatic reduction in the degree of Ni-O hybridization. This observation is in excellent conformance with the Ni$^{2+}$ charge state deduced from the Ni $L$-edge measurement.

Next we turn our attention to the question of how the electrical properties of ultra-thin LNO films are affected by the polar-misfit at the heterointerface.  Fig. 3(a) shows temperature-dependent resistivities for the series of samples. As clearly  seen,  15 uc and 10 uc films show bulk-like metallic behavior with a weak upturn at low temperature (inset of Fig. 3(a)). As the film thickness decreases, however, the resistivity increases and the 8 uc film remains metallic only down to 180 K; below this temperature the sample resistivity shows a semiconducting behavior. Consistent with the diffraction/absorption data discussed above, once the film thickness  cell approaches 6 uc, the sample exhibits insulating (semiconducting) behavior starting at the room temperature;  at the critical  thickness of 5 uc the film becomes highly insulating (2 orders of magnitude higher than the 6 uc film), thus signifying the development  of a new electronic ground state of the material.

While this metal-to-insulator transition as a function of film thickness could be linked to the effect of reduced dimensionality ~\cite{lno_jian_prb,lno_triscone}, the two orders of magnitude reduction in resistivity for the 5 uc LNO film grown on non-polar LAO (111)  and  the high temperature MIT at 250 K  point to the additional  effects responsible for such highly insulating behavior of 5 uc LNO film grown on STO (111). Again, the presence of a highly insulating LNO225 phase~\cite{res_la2ni2} can account for the strongly disparate behavior, thus  corroborating the conclusions deduced from the X-ray diffraction and XAS data. The details of these transport behavior are given in the Supplemental section.


Next we discuss  the movement of the additional XRD peak, marked by the black solid triangle in Fig. 2(a). In general, it is  assumed that oxygen vacancies in perovskite $AB$O$_{3-\delta}$ lattice are randomly arranged  and  characterized by the very  large  distribution of the local electronic environment~\cite{defect}. Based on this, the experimentally determined structure at any given temperature can be  viewed  as a spatial and time average of the different local structures. Among the  different possible structures with the formulae of $A_2B_2$O$_5$, La$_2$Ni$_2$O$_5$ adopts a very specific structure with alternating NiO$_6$ octrahedra and NiO$_4$ planar unit in the $ab$ plane (see top panel of Fig. 3(b)) due to the energetically unfavorable tetrahedral coordination of Ni$^{+2}$~\cite{defect,jpcb}.  At the  same time, the overall enhancement of electrical conductivity with increase in film thickness acts to reduce the polar catastrophe by partial {\em{metallic screening}}. Thus some of the oxygen vacancies in La$_2$Ni$_2$O$_5$ phase (formed during the first 5 uc growth)  can be further effectively compensated for by forming the La$_2$Ni$_2$O$_{5+\delta}$ phase  during the post annealing process under high oxygen pressure as shown schematically  in Fig. 3(b). This results in a shift of the XRD peak position towards   stoichiometric LaNiO$_3$. We can also conjecture that when the metallic screening is sufficiently strong, the  oxygen deficiencies are almost entirely compensated for as confirmed by the absence of the additional XRD peak for 15 uc film (Fig. 2(a)).

In order to understand how the oxygen vacancies screen the dipole field at the heterointerface, we  performed {\em ab-initio} calculations for STO/LNO slabs oriented along (111), with a 15-uc-thick STO substrate and LNO layers of two different thickness (3 uc and 9 uc). The results are shown in  Fig. 4.
The single oxygen vacancy formation energy $E_\textrm{form}$  is defined as  $E_\textrm{vac} - E_\textrm{full} + \mu_\textrm{O}\textrm{(}T, p_\textrm{O2}$),  where  $E_\textrm{vac}$  and $E_\textrm{full}$ is the free energy of a given slab with and without an oxygen vacancy respectively. The oxygen chemical potential $\mu_\textrm{O}$ ($T, p_\textrm{O2}$) depends on the temperature ($T$) and oxygen pressure ($p_\textrm{O2}$) during growth or post annealing and  can  be approximated as $E_\textrm{dis}$/2 ~\cite{codopedsto}, where $E_\textrm{dis} = -9.51$ eV  is  the dissociation energy to break up a single O$_\textrm{2}$ bond.
Fig.~4(b) shows the  oxygen vacancy formation energy ($E_\textrm{form}$) as a function of distance from the LNO/STO interface for 3 (blue) and 9 (red) LNO layers. As seen in Fig.~4(b) the formation energy is always smaller (and hence preferable) for the vacancy created over the LNO side (positive values along the horizontal axis) compared to that on the STO side (negative values on the horizontal axis) for both 15STO/3LNO and 15STO/9LNO systems. In addition, the calculation predicts a sharp decrease in $E_\textrm{form}$ upon approaching the surface monolayer of LNO (i.e. LNO/vacuum interface), and this  result is found to  be independent of the LNO slab  thickness. We note, however, that first-principles calculations do  not  take into account  the presence of an oxygen atmosphere during the growth and  post-annealing, which  acts to compensate the vacancies at the surface monolayer ~\cite{lao_sto_vacancy_prb}.   As  the oxygen ions are removed from a [La$O_3$]$^{-3}$ layer located deeper down from the LNO/vacuum interface, saturation of $E_\textrm{form}$  can be seen for the thicker LNO slab (9 LNO;  this saturation of $E_{form}$ is emphasized by the yellow area in Fig.~4(b)).  The shortest slab (3LNO) does not display such saturation due to the presence of the STO interface, where the formation energy continues to increase.

 Comparing the two curves in Fig. 4(b), it can be easily seen that    deep into the STO substrate 
 $E_\textrm{form}$ takes the same value for both of the slabs considered regardless of LNO thickness (as indicated by the `$\ast$' mark in Fig.~4(b)),  yet  formation of oxygen vacancies is always more favorable on LNO side for the 3LNO slab compared to 9 uc LNO. This corroborates the earlier discussed Ni $L_2$ edge XAS results, showing the presence of larger number of Ni$^{2+}$ ions in 5 uc film compared to the thicker film.

The electrostatic potential energies shown for 15STO/9LNO and 15STO/3LNO slabs in Figs.~4 (a) and (c) respectively, further aid in understanding how the polar catastrophe can be avoided with oxygen vacancies.  Those potentials are shown in black for systems without oxygen vacancies, and in green for slabs with a single oxygen vacancy  at locations highlighted by vertical arrows.   In addition,  the {\em mean} potential energy,   averaged over a the length of a unit cell  and displayed in red (blue) for slabs without oxygen vacancies (with a single oxygen vacancy),  allows the reader to better view the potential profile across these slabs. The important observation is that the mean potential energy has a slower gradient and hence a smaller electric field (proportional to the gradient of potential) across the interface for systems with oxygen vacancies formed at the vicinity of  the heterointerface (blue trendlines) as compared to the system with no vacancies (red trendlines).

We have also explored the effects of   {\em{oxygen di-vacancy}} on the potential energy profile across the STO/LNO system. Given a 15STO/9LNO system with a vacancy close to the heterointerface, the calculation reveals that the second oxygen vacancy has its lowest formation energy when the second oxygen is removed from the same initial NiO$_6$ octahedra but farthest away from the first vacancy site (shown by vertical arrows in Fig. 4(e)), thus forming a  square coordinated by oxygen Ni unit characteristic of the LNO225 phase. As immediately  seen in  Fig. 4(e), the creation of this 2$^{nd}$ vacancy results in an even smother slope of the average electrostatic potential (compare the smoother blue trend line vs. the red one). This  result thus affirms  a remarkably  efficient  suppression of the polar mismatch across the heterointerface and provides  further microscopic insight  in to the role of  oxygen vacancies as an efficient route for reducing, and even almost completely eliminating the polar mismatch at the oxides heterointerface.


To summarize,  a series of epitaxial ultra-thin LaNiO$_3$ films of different thicknesses (ranging  from 5-15 unit cell) have been grown on polar STO (111) and  non-polar LAO (111) substrates to investigate the effect of polar mismatch at the perovskite interface.  A  combination of $in-situ$ RHEED imaging recorded during the growth of each consecutive unit cell, X-ray diffraction, transport and synchrotron based resonant X-ray spectroscopy  at  Ni and O edges reveal the formation of polarity stabilized  chemical phase La$_2$Ni$_2$O$_5$ (Ni$^{2+}$) for a few unit-cell thick films. A gradual increase of metallicity with increasing thickness serves as an additional mechanism to screen the diverging potential, leading to a sharp decrease of oxygen vacancies for thicker films and restoring nominal 3+ charge state of Ni. The combined experimental and {\em ab-initio} results demonstrate  how the polarity at the oxide interface can be effectively controlled by  stabilizing another structural and electronic phase formed within the first few monolayers. These findings  should  be taken into consideration in addition to  electronic reconstruction and  quantum confinement when describing the properties of ultra-thin films of complex oxides with polar discontinuity.

\section*{Methods}
 {\bf Experimental Techniques:}   LaNiO$_3$ thin films with different thicknesses were grown on high-quality STO (111) and LAO (111) substrates (Crystec, Germany) by pulsed laser interval deposition~\cite{jian_apl,misha_jap,own_apl} (laser frequency: 18 Hz). In order to avoid additional surface defects on substrate, formed by the chemical treatment for achieving single termination~\cite{defects}, as received substrates were used.  50 mTorr  partial pressure of oxygen was maintained during the growth and all of the grown samples were subsequently post annealed in-situ for 30 min in 1 atm of ultra pure oxygen at growth temperature (670$^\circ$C), which was  found to be essential to maintain correct oxygen stoichiometry for the (001) oriented LNO/LAO heterostructures\cite{jian_apl}.   The films were characterized ex-situ by  laboratory-based XRD (Panalytical Xpert Pro MRD [Panalytical, Almelo]).  Ni $L_{3,2}$ edge and O $K$ edge XAS spectra were taken at room temperature at the 4-ID-C beam line of the Advanced Photon Source at Argonne National Laboratory. Electrical d.c. transport characterization was performed on a commercial physical properties measurement system (PPMS) with van der Paw geometry.
\

{\bf Theoretical Methods:}
LDA  calculations  on 15uc STO/$n-$uc LNO multilayer slabs oriented along the (111) direction were carried out with the Vienna ab initio simulation package (VASP)~\cite{vasp} using projector-augmented waves~\cite{paw,paw1}. Total energies have been calculated for the stoichiometric  structure with a 6x6x1 k-point sampling, with full structural relaxation and dipole corrections. The slabs have a 20 \AA\ - 25 \AA\  vacuum in both sides to minimize artificial interactions among the periodic images. In computing oxygen vacancies, total electronic energies were calculated by removing one oxygen (for each calculation) from different layers with full structural atomic relaxation and fixed lattice vectors.
 
\section*{Acknowledgement}

J. C. was supported by DOD-ARO under Grant No. 0402-17291 and DOE under Grant no. 0402 81814-21-0000. Work at the Advanced Photon Source, Argonne was supported by the U.S. Department of Energy, Office of Science under Grant No. DEAC02-06CH11357. S. M. thanks  M. Hawkridge for the help in XRD measurements and S. B.-L. thanks L. Bellaiche and H. Fufor discussions and funding from Arkansas Biosciences Institute. Calculations were carried out at TACC (Stampede, Grant XSEDE TG-PHY090002) and Razor (Arkansas).




\newpage

\begin{figure}[h!] 
\includegraphics[width=1\textwidth] {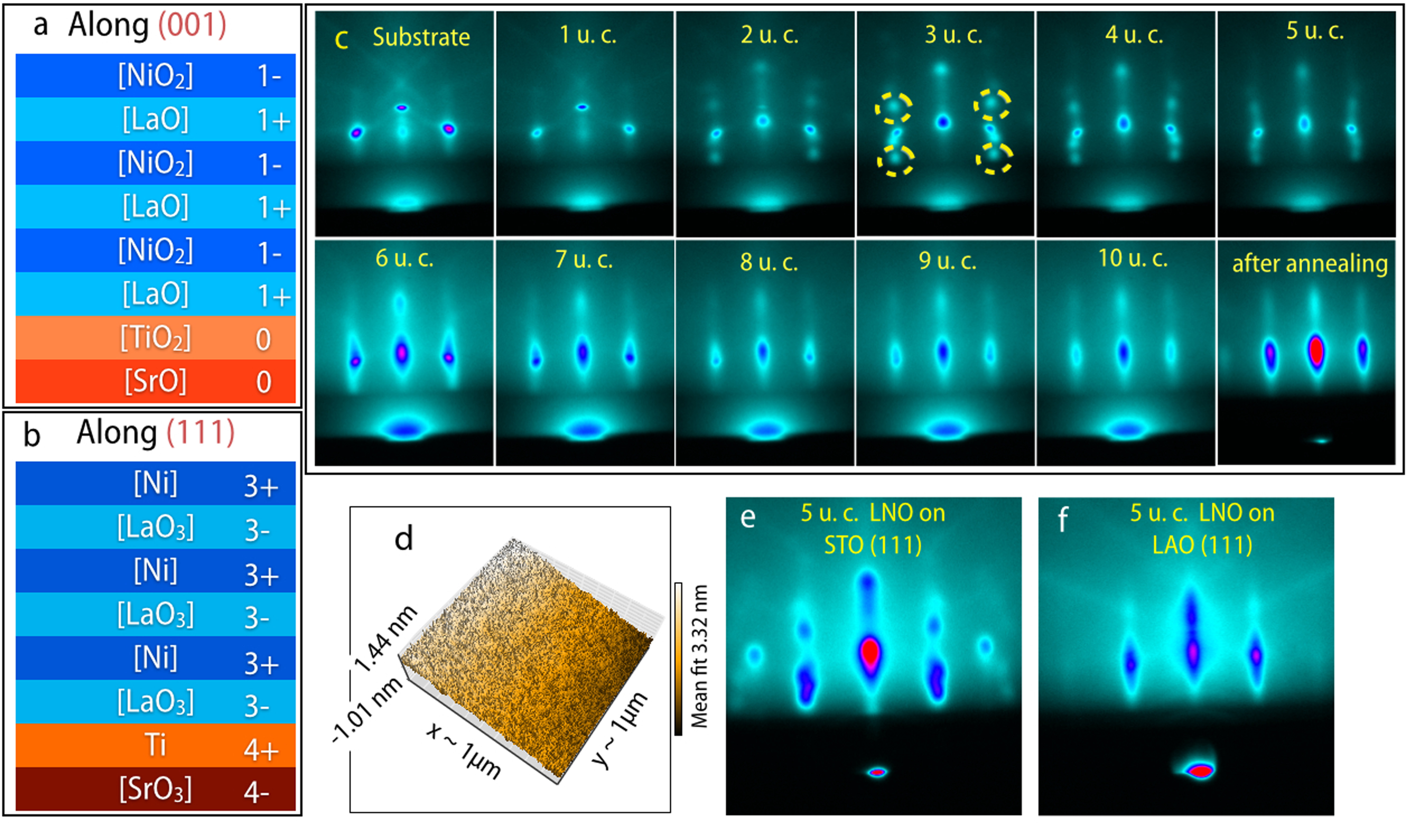}
\caption{  {\bf Polarity mismatch and morphology during growth.} (a) and (b) Schematics of polar discontinuity (in purely ionic limit and  without considering any surface reconstruction). STO (: SrO, TiO$_2$) is non-polar while LNO (: [LaO]$^{1+}$ [NiO$_2$]$^{1-}$) is polar  along the (001) direction . Both STO (: [SrO$_3$]$^{4-}$, Ti$^{4+}$) and  LNO (: [LaO$_3$]$^{3-}$Ni$^{3+}$) are strongly polar along the (111) direction. (c) Set of  RHEED images recorded during growth of 10 unit cell LNO on STO (111) substrate. RHEED patterns from bare STO (111) substrate and the film (after cooling to 300 K) are also shown.  The additional reflections  are highlighted by the dotted ellipses in one representative image for clarity. (d)  AFM image of a 10 uc LNO film. RHEED images for a 5 uc LNO film grown on (e) polar STO (111), (f) non-polar LAO (111) substrate. All  RHEED images are taken along the pseudo cubic [1-1 0] direction. All of the crystallographic directions, film thickness in this work are defined with respect to the  psedocubic unit cell.}
\end{figure}

\begin{figure}[h!]
\includegraphics[width=1\textwidth]{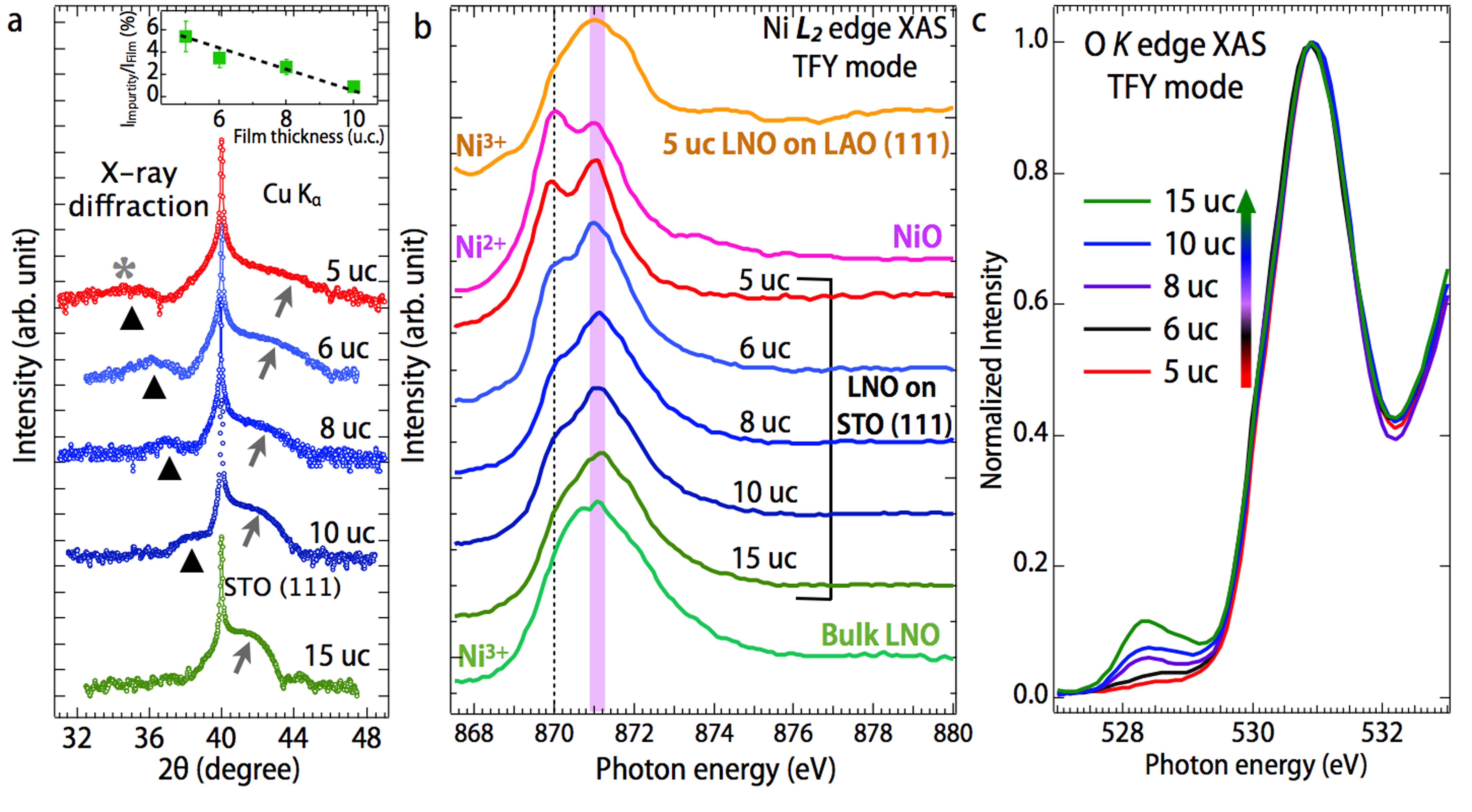}
\caption{\label{}  {\bf X-ray diffraction and X-ray absorption spectroscopy.}  (a) 2$\theta$ - $\omega$ scans for LNO films around the STO (111) reflection. The (111) film peaks and the `impurity' peaks are indicated by arrow and solid triangle respectively.  The ratio of the area under the `impurity'  peak and film peak is shown as a function of film thickness in the  inset. (b)   Ni $L_2$-edge and (c) O $K$-edge XAS spectra  from different samples recorded in TFY mode at 300 K. }
\end{figure}

\begin{figure}[h!] 
\includegraphics[width=.70\textwidth]{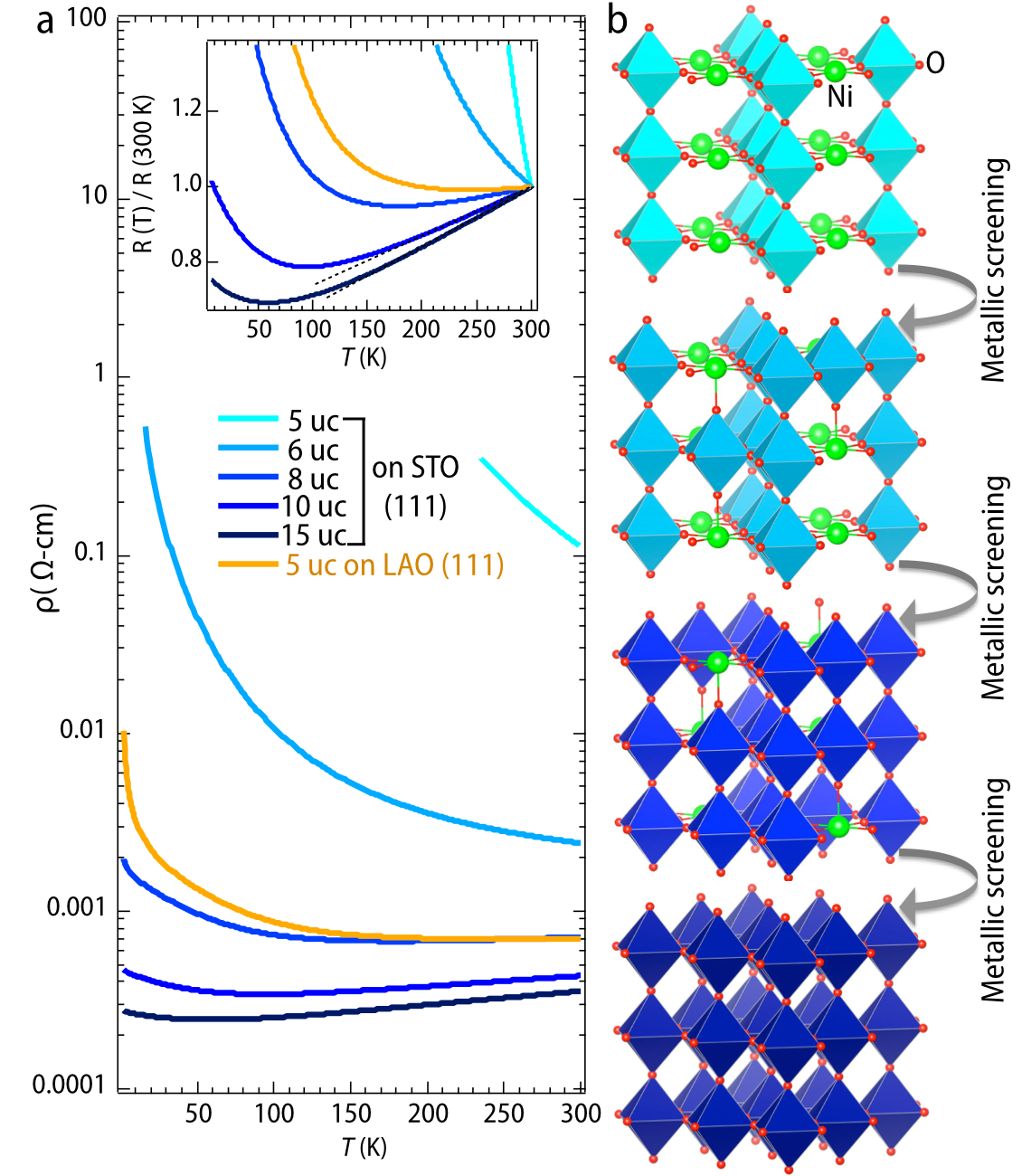}
\caption{\label{} {\bf Transport properties.} (a) Resistivity  as a function on temperature for LNO films on STO (111) with different thickness.  Inset shows the variation of scaled resistivity. The dotted lines  for 15 and 10 uc film there indicate linear variation of $\rho$ with $T$ in metallic region.  The enhancement of  electrical conductivity with the film thickness partially screens the interfacial field, resulting  in a gradual compensation of oxygen vacancies in La$_2$Ni$_2$O$_5$ phase (formed during the first 5 uc growth). Such evolution of La$_2$Ni$_2$O$_{5+\delta}$ structure at the film-substrate interface has been shown schematically in (b). The color of the NiO$_6$ octahedra  represents the sample thickness mentioned in (a) by the same color of $\rho$ vs. $T$ graph. The peak at 531 eV is due to the STO substrate. }
\end{figure}


\begin{figure}[h!]
\includegraphics[width=1\textwidth]{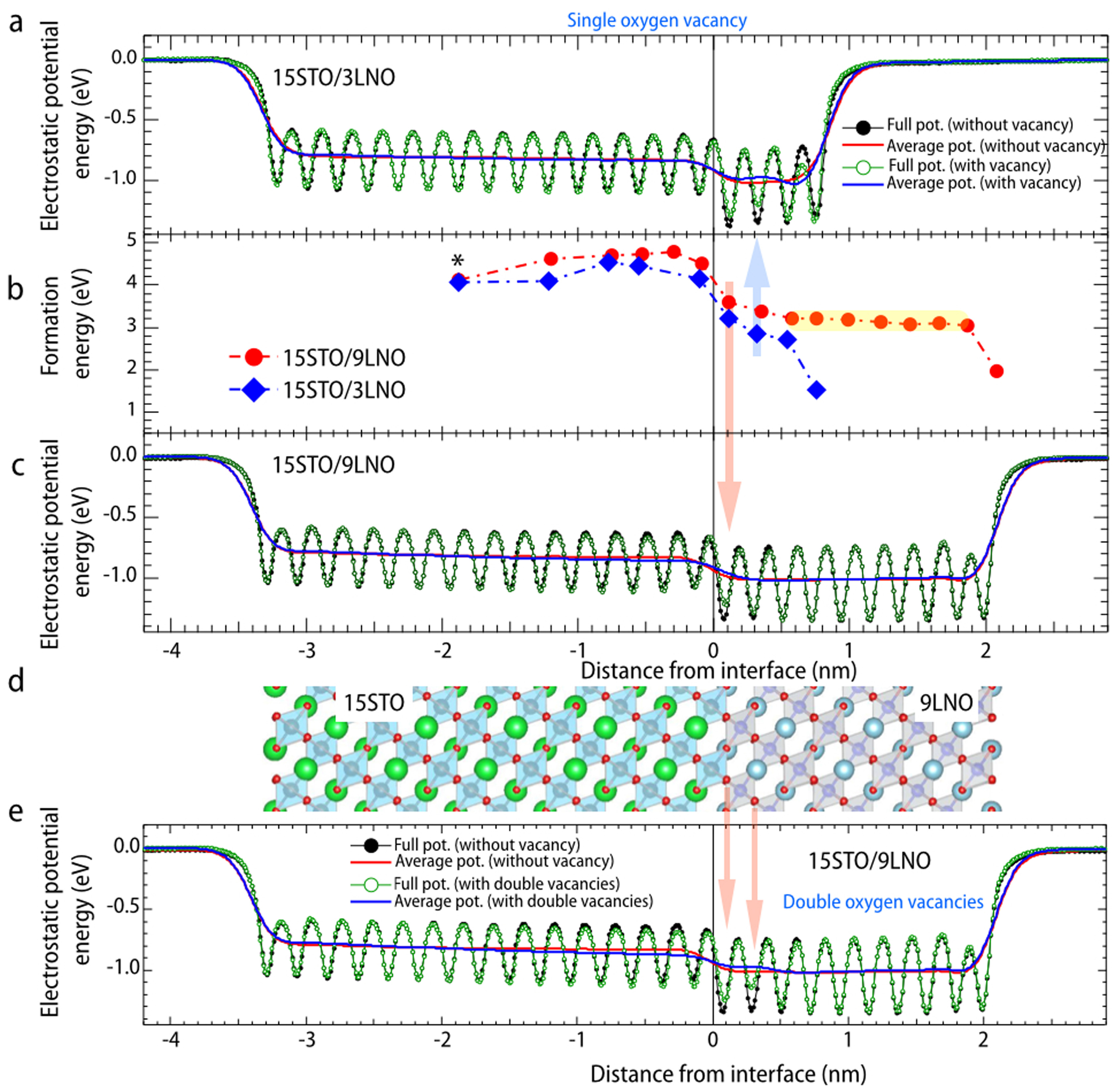}
\caption{\label{}   {\bf Potential energy and oxygen-vacancy formation energy from first principles.}    (a and c) The electrostatic potential for 15STO/9LNO and 15STO/3LNO slabs with and without oxygen vacancy (the location of the vacancy has been shown by vertical arrow for both cases). (b) Oxygen-vacancy formation energy along the LNO/STO slab. Each of the data point corresponds to a calculation with a single vacancy at that location. (d) 15STO/9LNO slab used in calculations. (e) The electrostatic potential for 15STO/9LNO with two oxygen vacancies (positions of the vacancies are indicated by arrow in (d)).}
\end{figure}


\begin{thebibliography}{99}

\bibitem{jak_nm} Chakhalian, J.,  Millis, A. J. \&  Rondinelli, J. Whither the oxide interface. $Nature \ Mater.$ {\bf 11,} 92-94 (2012).
\bibitem{tokura_nm_12} Hwang, H. Y. $et \ al.$  Emergent phenomena at oxide interfaces. $Nature \ Mater.$ {\bf 11,} 103-113 (2012).
\bibitem{ramesh_mat} Yu, P.,  Chu, Y.-H. \&  Ramesh, R. Oxide interfaces: pathways to novel phenomena. $Mater. \ Today$ {\bf 15,} 320-327 (2012).
\bibitem{bibes} Bibes M.,  Villegas Javier E.  \& Barth$\acute{e}$l$\acute{e}$my A., Ultrathin oxide films and interfaces for electronics and spintronics. $Advances \ In \ Physics$ {\bf 60,}  5-84 (2010).
\bibitem{lao_nature}  Ohtomo, A. \& . Hwang, H. Y. A high-mobility electron gas at the LaAlO$_3$/SrTiO$_3$ heterointerface. $Nature$  {\bf 427,} 423-426 (2004).
\bibitem{lao_science} Reyren, N. $et \ al.$ Superconducting Interfaces Between Insulating Oxides, $Science$ {\bf 317,} 1196-1199 (2007).


\bibitem{lao_nm}  Nakagawa, N.,   Hwang, H. Y. \&   Muller, D. A. Why some interfaces cannot be sharp. {\it Nat. Matt.} {\bf 5,} 204-209 (2006).
\bibitem{lao_sto_opt}  Savoia, A. $et \ al.$ Polar catastrophe and electronic reconstructions at the LaAlO$_3$/SrTiO$_3$ interface:  Evidence from optical second harmonic generation. {\it Phys. Rev. B} {\bf 80,} 075110 (2009).
 \bibitem{lao_sto_hardxps} Sing, M. $et \ al.$  Profiling the Interface Electron Gas of LaAlO$_3$/SrTiO$_3$ Heterostructures with Hard X-Ray Photoelectron Spectroscopy. {\it Phys. Rev. Lett.} {\bf 102,} 176805 (2009).
\bibitem{lao_sto_mixing}  Salluzzo, M. $et \ al.$ Structural and Electronic Reconstructions at the LaAlO$_3$/SrTiO$_3$ Interface. {\it Adv. Mater.}   {\bf 25,} 2333  (2013).
\bibitem{lao_sto_vacancy1} Herranz, G.  $et \ al.$ High Mobility in LaAlO$_3$/SrTiO$_3$ Heterostructures: Origin, Dimensionality, and Perspectives. {\it Phys. Rev. Lett.}  {\bf 98,} 216803 (2007).
\bibitem{lao_sto_vacancy} Park, J. $et \ al.$ Oxygen-Vacancy-Induced Orbital Reconstruction of Ti Ions at the Interface of LaAlO$_3$/SrTiO$_3$ Heterostructures: A Resonant Soft-X-Ray Scattering Study. {\it Phys Rev. Lett.} {\bf 110,} 017401 (2013).
\bibitem{lao_admi} Lin, W.-N. $et \ al.$ Electrostatic Modulation of LaAlO$_3$ /SrTiO$_3$ Interface Transport in an Electric Double-Layer Transistor. {\it Adv. Mater. Interfaces} {\bf 1}, 1300001 (2014).

\bibitem{llvo_sto} Hotta, Y.,  Susaki, T. \&  Hwang, H. Y. Polar Discontinuity Doping of the LaVO$_3$/SrTiO$_3$ Interface. {\it Phys. Rev. Lett.} {\bf 99,} 236805 (2007).
\bibitem{lao_lvo} Takizawa, M.  $et \ al.$  Spectroscopic evidence of competing interactions in polar multilayers LaAlO$_3$/LaVO$_3$/LaAlO$_3$. {\it Phys. Rev. Lett.} {\bf 102,} 236401 (2009).
\bibitem{lco_sto}  Chambers, S. A. $et \ al.$ Band allignment, built-In Potential, and the absence of conductivity at the LaCrO$_3$/SrTiO$_3$ (001) heterojunction.  {\it Phys. Rev. Lett.} {\bf 107,} 206802 (2011).
\bibitem{lufeo3}Akbashev, A.R. $et \ al.$ Reconstruction of the polar interface between hexagonal LuFeO$_3$ and intergrown Fe$_3$O$_4$ nanolayers. {\it Sci. Rep.} {\bf 2,} 672 (2012).
\bibitem{lao_sto_vacancy_prb}  Zhong, Z.,  Xu, P. X. \&  Kelly, P. J. Polarity-induced oxygen vacancies at LaAlO$_3$/SrTiO$_3$ interfaces. {\it Phys. Rev. B} {\bf 82,} 165127 (2010).

\bibitem{lno1_fiete}R\"{u}egg, A.  \& Fiete, G. A. Topological insulators from complex orbital order in transition-metal oxides heterostructures. {\it Phys. Rev. B} {\bf 84,} 201103(R) (2011).
\bibitem{satoshi_prb}  Yang, K. -Y. $et \ al.$ Possible interaction-driven topological phases in (111) bilayers of LaNiO$_3$. {\it Phys. Rev. B} {\bf 84,} 201104(R) (2011).
\bibitem{sriro_prb}  Wang, F.  \&  Ran, Y. Nearly flat band with Chern number $C$=2 on the dice lattice. {\it Phys. Rev. B} {\bf 84,} 241103(R) (2011).
\bibitem{nagaosa_nc} Xiao, D.  $et \ al.$  Interface engineering of quantum Hall effects in digital transition metal oxide heterostructures. {\it Nat. Commun.} {\bf 2,} 596 (2011).
\bibitem{lno_fiete} R\"{u}egg, A. $et \ al.$ Electronic structure of (LaNiO$_3$)$_2$/(LaAlO$_3$)$_N$ heterostructures grown along [111]. {\it Phys. Rev. B} {\bf 85,} 245131 (2012).
\bibitem{okamoto_prl} Okamoto, S.  Doped Mott Insulators in (111) Bilayers of Perovskite Transition-Metal Oxides with a Strong Spin-Orbit Coupling. {\it Phys. Rev. Lett.} {\bf 110,} 066403 (2013).
\bibitem{lno_strain_fiete}R\"{u}egg, A. $et \ al.$   Lattice distortion effects on topological phases in LaNiO$_3$)$_2$/(LaAlO$_3$)$_N$ heterostructures grown along the [111] direction. {\it Phys. Rev. B} {\bf 88,} 115146 (2013).



\bibitem{blok}Blok J. L.,  $et \ al.$ Epitaxial oxide growth on polar (111) surfaces.  {\it Appl. Phys. Lett.} {\bf 99,} 151917 (2011).
\bibitem{jian_apl}  Liu, J. $et \ al.$ Effect of polar discontinuity on the growth of LaNiO$_3$/LaAlO$_3$ superlattices.  {\it Appl. Phys. Lett.} {\bf 96,} 133111 (2010).
\bibitem{lno_jian_prb}Liu, J. $et \ al.$ Quantum confinement of Mott electrons in ultrathin LaNiO$_3$/LaAlO$_3$ superlattices. {\it Phys. Rev. B} {\bf 83,} 161102(R) (2011).
\bibitem{lno_epl}  Freeland, J. W. $et \ al.$ Orbital control in strained ultra-thin LaNiO$_3$/LaAlO$_3$ superlattices. {\it Euro Phys. Lett.} {\bf 96,} 57004 (2011).
\bibitem{lno_keimer_science} Boris, A. V. $et \ al.$ Dimensionality Control of Electronic Phase Transitions in Nickel-Oxide Superlattices.  {\it Science} {\bf 332,} 937-940 (2011).
\bibitem{lno_keimer_nm} Benckiser, E. $et \ al.$ Orbital reflectometry of oxide heterostructures. {\it Nat. Mater.} {\bf 10,}  189-193 (2011).
\bibitem{lno_triscone}Scherwitzl, R. $et \ al.$ Metal-insulator transition in ultrathin LaNiO$_3$ films. {\it Phys. Rev. Lett.} {\bf 106,} 246403 (2011).
\bibitem{jak_prl} Chakhalian J. $et \ al.$ Asymmetric Orbital-Lattice Interactions in Ultrathin Correlated Oxide Films. {\it Phys. Rev. Lett.}  {\bf 107,} 116805 (2011).
\bibitem{lno_th1}Chaloupka, J.  \&  Khaliullin, G. Orbital Order and Possible Superconductivity in LaNiO$_3$/LaMO$_3$ Superlattices. {\it Phys. Rev. Lett.} {\bf 100,} 016404 (2008).
\bibitem{lno_th2} Hansmann, P. $et \ al.$ Turning a Nickelate Fermi Surface into a Cupratelike One through Heterostructuring.  {\it Phys. Rev. Lett.} {\bf 103,} 016401 (2009).



\bibitem{sto111}R\"{o}del, T. C.  $et \ al.$ Orientational Tuning of the Fermi Sea of Confined Electrons
at the SrTiO$_3$ (110) and (111) Surfaces, {\it Phys. Rev. Appl.} {\bf 1}, 051002 (2014).
\bibitem{rheed1} C. P. Wang S. $et \ al.$ Deposition of in-plane textured MgO on amorphous Si$_3$N$_4$ substrates by ion-beam-assisted deposition and comparisons with ion-beam-assisted deposited yttria-stabilized-zirconia, {\it Appl. Phys. Lett.} {\bf 71,} 2955 (1997).
\bibitem{misha_jap} Kareev, M. $et \ al.$  	Sub-monolayer nucleation and growth of complex oxides at high supersaturation and rapid flux modulation.   {\it J. Appl. Phys.} {\bf 109,} 114303 (2011).
\bibitem{edoped} Middey,  S. $et \ al.$ Epitaxial stabilization of ultra thin films of electron doped manganites. {\it Appl. Phys. Lett.} {\bf 104}, 202409 (2014).
\bibitem{own_apl} Middey,  S. $et \ al.$ Epitaxial growth of (111)-oriented LaAlO$_3$/LaNiO$_3$ ultra-thin superlattices. {\it Appl. Phys. Lett.} {\bf 101,} 261602 (2012).




\bibitem{la2ni2}   Alonso, J. A. \&  Martinezlope, M. J. Preparation and crystal-structure of the deficient perovskite LaNiO$_{2.5}$, solved from neutron diffraction data. {\it J. Chem. Soc., Dalton Trans.} {\bf 1995,} 2819-2824 (1995).
\bibitem{lani}  Crespin, M. Levitz, P.  \& Gatineau, L.   Reduced forms of LaNiO$_3$ perovskite  .1. evidence for new phases - La$_2$Ni$_2$O$_5$ and LaNiO$_2$. {\it  J. Chem. Soc. Faraday Trans.}  {\bf 79,} 1181 (1983).


\bibitem{oxygen_kedge}Abbate, M. $et \ al.$ Electronic structure and metal-insulator transition in LaNiO$_{3-\delta}$. {\it Phys. Rev. B} {\bf 65,} 155101 (2002).
\bibitem{res_la2ni2}S$\acute{a}$nchez, R. D.  $et \ al.$  Metal-insulator transition in oxygen-deficient LaNiO$_{3-x}$ perovskites. {\it Phys. Rev. B} {\bf 54,} 16574 (1996).
\bibitem{defect} St${\o}$len, S. Bakken, E.  \& Mohn, C. E. Oxygen-deficient perovskites: linking structure, energetics and ion-transport. {\it  Phys. Chem. Chem. Phys.} {\bf 8,} 429 (2006).
\bibitem{jpcb} St${\o}$len, S. Mohn, C. E. Ravundran, P.  \& Allan, N. L. Topography of the Potential Energy Hypersurface and Criteria for Fast-Ion Conduction in
Perovskite-Related A$_2$B$_2$O$_5$ Oxides. {\it J. Phys. Chem. B} {\bf 109,} 13262 (2005).
\bibitem{codopedsto} Posadas A. B. $et \ al.$ Oxygen vacancy-mediated room-temperature ferromagnetism in insulating cobalt-substituted SrTiO$_3$ epitaxially integrated with silicon {\it Phys. Rev. B} {\bf 87},  144422 (2013).

\bibitem{defects}  Zhang, J.  $et \ al.$ Depth-resolved subsurface defects in chemically etched SrTiO$_3$, {\it Appl. Phys. Lett.} {\bf 94}, 092904 (2009).
\bibitem{vasp} Kresse G. \&   Furthm$\ddot{u}$ller J. Efficient iterative schemes for ab initio total-energy calculations using a plane-wave basis set. {\it Phys. Rev. B} {\bf 54}, 11169 (1996).
\bibitem{paw} Bl$\ddot{o}$chl P. E. Projector augmented-wave method. {\it Phys. Rev. B} {\bf 50}, 17953 (1994).
\bibitem{paw1} Kresse G. \& Joubert D.  From ultrasoft pseudopotentials to the projector augmented-wave method. {\it Phys. Rev. B} {\bf 59}, 1758 (1999).


\end{thebibliography}
 \end{document}